\documentclass[twocolumn]{aastex7}

\newcommand\wsa{\texttt{WSA}}
\newcommand\euhforia{\texttt{EUHFORIA}}
\newcommand\huxt{\texttt{HUXt}}
\newcommand\mas{\texttt{MAS}}
\newcommand\gtm{“ground truth map”}

\newcommand{\spwea}{Space Weather}
\newcommand{\jswsc}{J. Space Weather Space Clim.}

\usepackage{comment}
\usepackage{amsmath}
\usepackage{multirow}
\usepackage{booktabs}
\usepackage{hyperref}
\shorttitle{ .. }
\shortauthors{Heinemann et al.}

\begin{document}

\title{Quantifying Uncertainties in Solar Wind Forecasting Due to Incomplete Solar Magnetic Field Information}

\author[0000-0002-2655-2108]{Stephan~G.~Heinemann}
\affiliation{Department of Physics, University of Helsinki, P.O. Box 64, 00014, Helsinki, Finland}
\email{stephan.heinemann@hmail.at}

\author[0000-0003-1175-7124]{Jens~Pomoell}
\affiliation{Department of Physics, University of Helsinki, P.O. Box 64, 00014, Helsinki, Finland}
\email{jens.pomoell@helsinki.fi}

\author[0000-0002-2633-4290]{Ronald~M.~Caplan}
\affiliation{Predictive Science Inc., 9990 Mesa Rim Road, Suite 170, San Diego, CA 92121, USA}
\email{caplanr@predsci.com}

\author[0000-0003-2061-2453]{Mathew J. Owens}
\affiliation{Space and Atmospheric Electricity Group, Department of Meteorology, University of Reading, Earley Gate, P.O. Box 243, Reading RG6 6BB, UK}
\email{m.j.owens@reading.ac.uk}

\author[0000-0001-9498-460X]{Shaela~Jones}
\affiliation{Heliophysics Science Division, NASA Goddard Space Flight Center, Code 671, Greenbelt, MD,20771, USA}
\email{shaela.i.jonesmecholsky@nasa.gov}

\author[0000-0003-0621-4803]{Lisa~Upton}
\affiliation{Southwest Research Institute, Boulder, CO 80302, USA}
\email{lisa.upton@swri.org}

\author[0000-0003-3191-4625]{Bibhuti~Kumar~Jha}
\affiliation{Southwest Research Institute, Boulder, CO 80302, USA}
\email{bibhuti.jha@swri.org}

\author[0000-0001-9326-3448]{Charles N. Arge}
\affiliation{Heliophysics Science Division, NASA Goddard Space Flight Center, Code 671, Greenbelt, MD,20771, USA}
\email{charles.n.arge@nasa.gov}

\begin{abstract}
Solar wind forecasting plays a crucial role in space weather prediction, yet significant uncertainties persist due to incomplete magnetic field observations of the Sun. Isolating the solar wind forecasting errors due to these effects is difficult. This study investigates the uncertainties in solar wind models arising from these limitations. We simulate magnetic field maps with known uncertainties, including far-side and polar field variations, as well as resolution and sensitivity limitations. These maps serve as input for three solar wind models: the Wang-Sheeley-Arge (WSA), the Heliospheric Upwind eXtrapolation (HUXt), and the European Heliospheric FORecasting Information Asset (EUHFORIA). We analyze the discrepancies in solar wind forecasts, particularly the solar wind speed at Earth's location, by comparing the results of these models to a created “ground truth” magnetic field map, which is derived from a synthetic solar rotation evolution using the Advective Flux Transport (AFT) model. The results reveal significant variations within each model with a RMSE ranging from $59-121~\mathrm{km~s^{-1}}$. Further comparison with the thermodynamic Magnetohydrodynamic Algorithm outside a Sphere (MAS) model indicates that uncertainties in the magnetic field data can lead to even larger variations in solar wind forecasts compared to those within a single model. However, predicting a range of solar wind velocities based on a cloud of points around Earth can help mitigate uncertainties by up to $20-77\%$.

\end{abstract}

\keywords{Solar Wind (1534) --- Solar magnetic fields(1503) --- Space weather(2037) --- Heliosphere(711)}

\section{Introduction} \label{sec:intro}
Space weather has become an integral part of modern life, influencing various aspects of daily activities. Any space weather forecast — whether for eruptive or recurrent events — relies on the assumption of a fundamentally accurate (zeroth-order) ambient solar wind. Consequently, solar wind forecasting, and especially solar wind modeling, must begin with this foundation as highlighted by \cite{Case2008}.\\% \textbf{Large uncertainties in forecasting solar ejecta \citep[\textit{e.g., coronal mass ejections, or CMEs}][]{Amerstorfer2018} can be significantly reduced by incorporating a more accurate representation of the solar wind.} \\

To forecast space weather impacts at a high level, the ambient solar wind must be reliably understood, modeled, and predicted. Currently, a wide range of methods is employed for solar wind forecasting, including data-driven, empirical and semi-empirical approaches \citep[\textit{e.g.,}][]{2012rotter,2015rotter,Reissetal2016,Temmer2018,milosic2023}, hybrid and physics-based models \citep[see][]{Owens2008}, and numerical modeling approaches ranging from 1D to 3D simulations \citep[][]{Pomoell2019,OdstrcilPizzo1999,Mikic99,Barnard2022}. However, studies have highlighted significant uncertainties even in processes considered among the simplest in space weather forecasting—the ambient solar wind\citep[\textit{e.g.,}][]{Jian2015,Hinterreiter2019,Reiss2023}. \\

Among all potential sources of uncertainty, the lack of concurrent observational coverage of the Earth-Sun system has been identified as a primary challenge \citep[COSPAR ISWAT Roadmap,][]{Temmer2023}. This is because most models depend on a representation of the full-Sun surface radial magnetic field as input for forecasting the solar wind. However, the limited vantage points currently available routinely—primarily Earth—result in magnetic field maps that include inherent assumptions about the strength and distribution of the magnetic field, particularly in poorly observed regions such as the polar areas \citep[][]{Riley2019} and the solar far side \citep[\textit{e.g.,}][]{Heinemann2021_farside,Yang2024_FARM}.\\

To mitigate these limitations, two common approaches are widely used in the community. The first involves the creation of synoptic charts, which are constructed by effectively stacking bands of magnetic field observations near the central meridian over the course of a solar rotation. However, this method neglects the potential rapid evolution of the solar magnetic field, leading to an “aging effect” \citep[][]{Heinemann2021_farside, Posner2021}, where some parts of the map are significantly older than others. The second approach uses surface transport models to evolve the solar surface magnetic field observed from Earth, based on theoretical assumptions \citep[\textit{e.g.,}][]{Argeetal2010,AFT2014,Yang2024_FARM,Caplan2025_hipft}. While these models account for magnetic flux transport processes, they still lack observational data from unobserved regions of the Sun, where flux emergence and reconnection processes occur.\\

The most straightforward solution to these issues would be to directly observe the unmonitored regions of the Sun. Currently, the only instrument capable of partially observing the Sun's far side is the Photospheric and Magnetic Imager \citep[PHI;][]{2020solanki_so_phi} aboard {Solar Orbiter} \citep[SO;][]{2020muller_solO}. \citet{PHI_impact_2024} demonstrated that missing even a single far-side active region can significantly impact both local and global magnetic structures and \cite{downs24e} used PHI data to improve eclipse simulations. However, due to the complexity of SO's orbit, comprehensive far-side coverage is rare, and not available in real time. \\

\begin{figure*}
\centering \includegraphics[width=1\linewidth,angle=0]{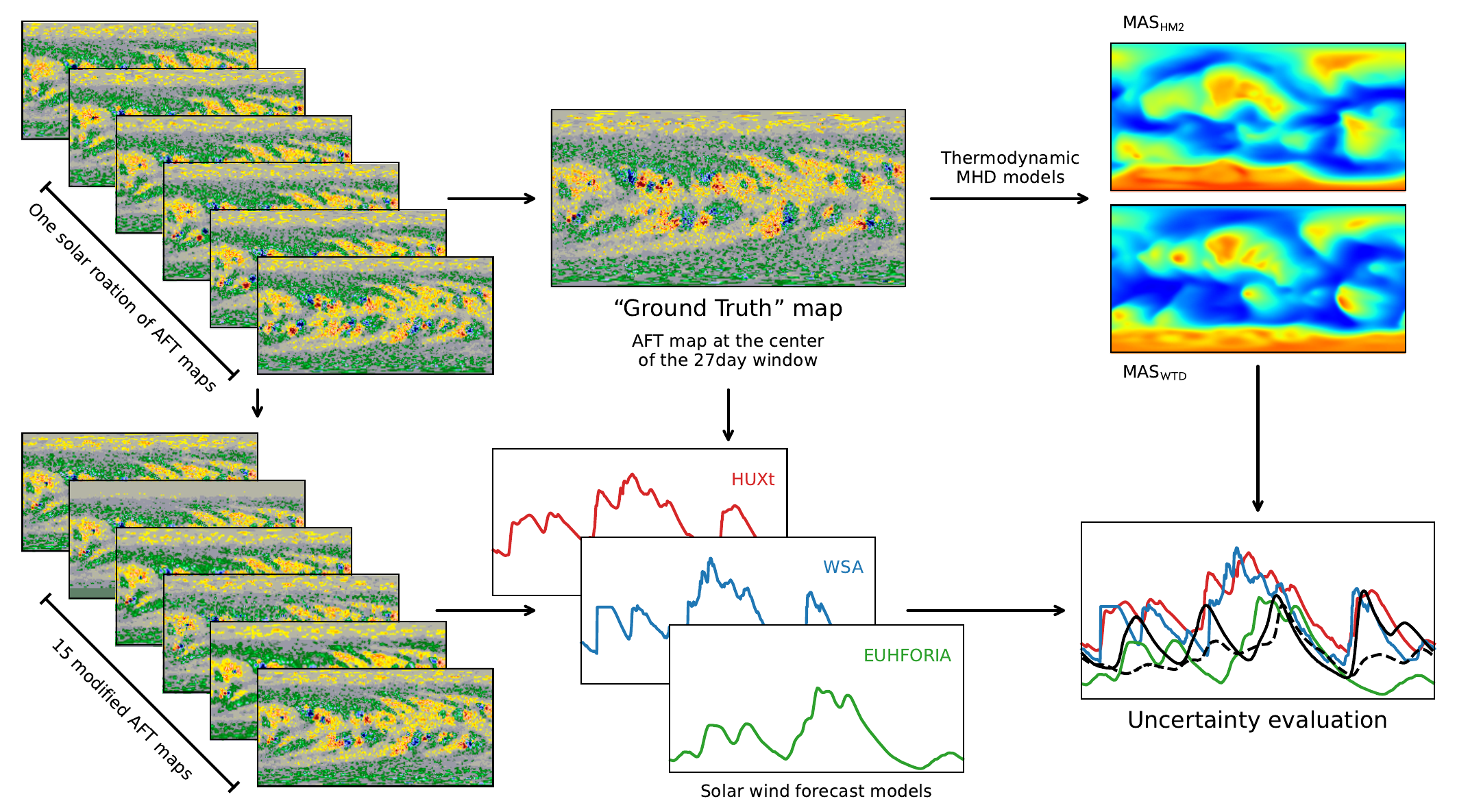}
\caption{Schematic of the methodological approach used in this study. We begin by analyzing one solar rotation of artificial magnetic fields to determine the \gtm\ and generate modified maps that account for observational uncertainties in 15 different ways. Next, we run three operational models with the modified maps: \huxt, \wsa, and \euhforia, and compare their results to those obtained from the \gtm, as well as from the \mas\ simulations of the \gtm.}\label{fig:schematic}
\end{figure*}

In this study, we investigate the uncertainties in solar wind forecasting arising from incomplete magnetic field information. To achieve this, we use a series of artificial magnetograms \citep[similar to the one used in][]{Jha2024} designed to emulate the most prominent uncertainties present in magnetic charts commonly employed in space weather modeling. By inputting these magnetograms into three operational solar wind models, we quantify the uncertainties stemming from these specific sources over one artificial solar rotation.  A general overview of the methodological approach of this study is shown in Figure~\ref{fig:schematic}.

\section{Data} \label{sec:data}
To quantify uncertainties from a specific origin, a ground truth must be established. In our case, this would ideally involve concurrent observations of the full solar magnetic field, which is currently not feasible. Therefore, we opted to use an artificial ground truth. Specifically, we simulate the magnetic field evolution over one solar rotation using the Advective Flux Transport model \citep[AFT;][]{AFT2014}. AFT has the ability to incorporate magnetic flux directly through data assimilation of magnetograms or by adding idealized bipoles to simulate active region emergence. In this case, we run AFT without data assimilation, and instead add synthetic active regions. Details of the model setup are provided in Section~\ref{sec:aft}. This approach yields 27 synthetic full-Sun magnetic field maps at a daily cadence with a resolution of $512\times1024$ pixels. \\

For this study, we assume that this set of maps represents the actual state of the Sun, \textit{i.e.,} the “ground truth”. From this set, we define the central map (day 13) as the \gtm\ ({Figure}~\ref{fig:maps1}a). For this map, we can apply pre-processing routines to make it compatible as input for thermodynamic magnetohydrodynamic (MHD) modeling (for details, see Section~\ref{sec:mas}; {Figure}~\ref{fig:maps1}b).

Using this set of observations, we derive maps that simulate known uncertainties arising from incomplete observations of the full Sun. Broadly, these uncertainties can be categorized into three main types:\\

\textit{Far-side uncertainty:} The standard method for creating solar maps for space weather modeling involves the use of synoptic charts. Synoptic charts are created by stacking overlapping $15^{\circ}$ slices taken from the central meridian to produce a map of the entire Sun that includes active region at all longitudes (see e.g., {Figure}~\ref{fig:maps1}c). Here, we have used all 27 AFT maps to create a synoptic chart that is constructed in the same manner as one created from daily observations. This approach mimics Earth as a single observation point. \\

In order to emulate the uncertainty due to the decreasing availability of magnetic data from Earth's vantage (as the flux rotates across the far-side without being updated), we apply a Gaussian filter to the \gtm\ to create a synthetic surface flux transport model-like map.  The Gaussian filter has a kernel size that increases as a function of distance from the assumed Earth-facing location (i.e., the most recent map) in direction of the solar rotation. We note that this smoothing is particularly evident on smaller spatial scales, over time ({Figure}~\ref{fig:maps1}g).\\ 

\textit{Uncertainty of the polar fields:} It is well known that polar fields are integral to heliospheric models, and their uncertainties may play a key role in addressing the issue of the missing open solar magnetic flux \citep[\textit{e.g.,}][]{Riley2019}. To simulate this behavior, we increase and decrease the polar magnetic flux of the synoptic chart by $\pm30\%$ using a \texttt{sin\textsuperscript{6}} function ({Figure}~\ref{fig:maps2}i,j). Additionally, we replicate the methodology used for the HMI pole-filled maps. Specifically, we replace the magnetic field above $68.5^{\circ}$ latitude uniformly with the mean value ({Figure}~\ref{fig:maps1}h).\\

\textit{Uncertainty due to resolution and sensitivity of the instrument:} Instrument-related uncertainties are highly specific and, therefore, difficult to assess accurately. Instead, we chose to simulate common pre-processing adjustments applied to magnetic maps for modeling purposes, specifically changes in resolution and smoothing. To do this, we doubled, halved, and quartered the pixel resolution of the synoptic charts while ensuring flux conservation ({Figure}~\ref{fig:maps1}d-f). We also applied various Gaussian filters uniformly across the entire synoptic chart, using six kernels with sizes of 0.5, 1.0, 1.5, 2.0, 2.5, and 3.0 degrees to remove different scales of information ({Figure}~\ref{fig:maps2}k-p).\\

In total, this leads to 16 different maps (including the \gtm) that we used as input to different solar wind models.

\begin{figure*}
\centering \includegraphics[width=1\linewidth,angle=0]{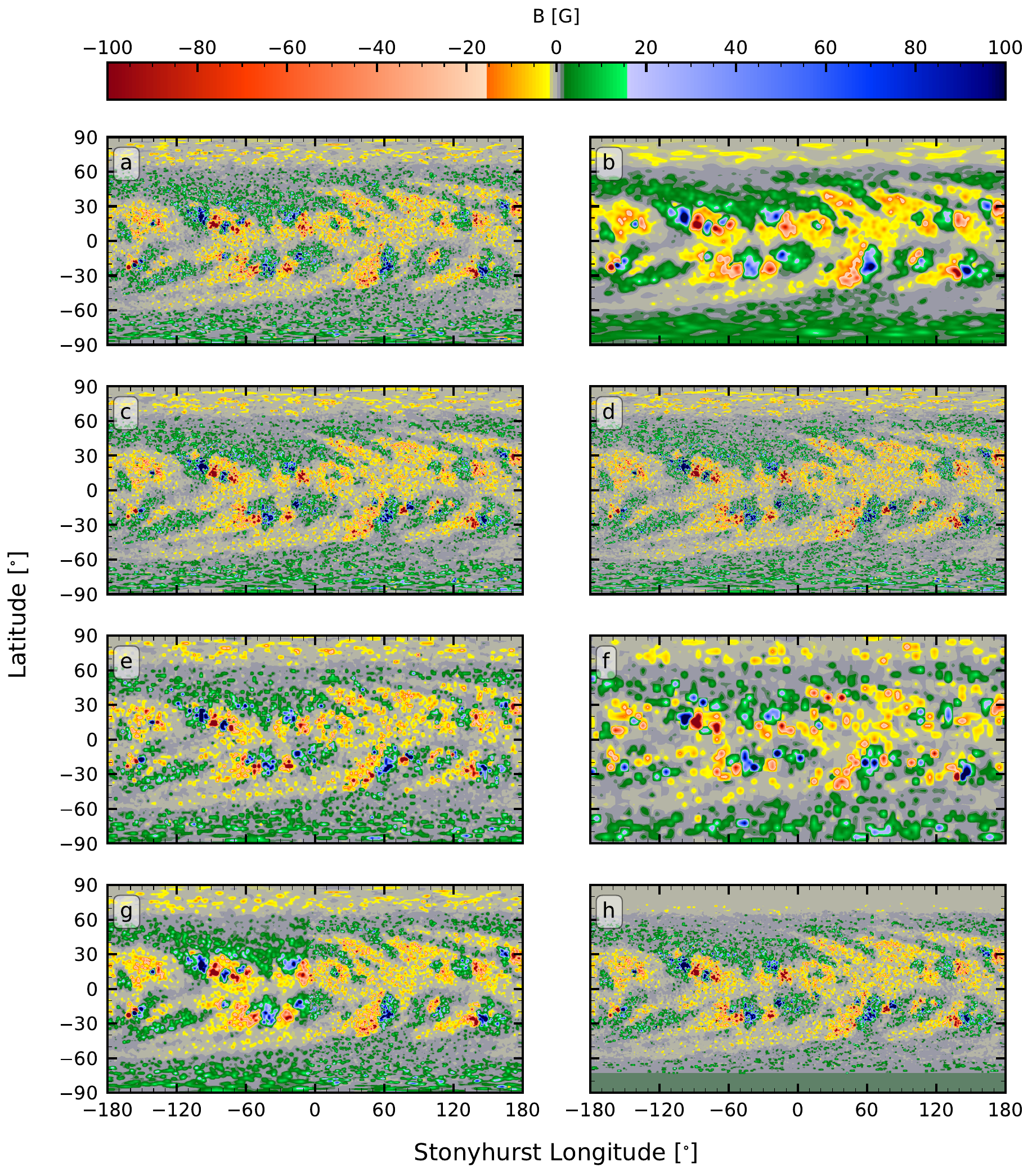}
\caption{Magnetic maps used in this study. (a) The base or "ground truth" map. (b) The "ground truth" map modified for \mas\  model runs. (c) A synoptic chart created from a 27-day sample of maps. (d-f) Synoptic charts with varying resolutions: double, half, and quarter, respectively. (g) The "ground truth" map with increased smoothing as a function of distance from the Earth-facing point, simulating uncertainty in surface flux transport models. (h) A synoptic chart with uniform field strength applied in the polar regions.}\label{fig:maps1}
\end{figure*}
\begin{figure*}
\centering \includegraphics[width=1\linewidth,angle=0]{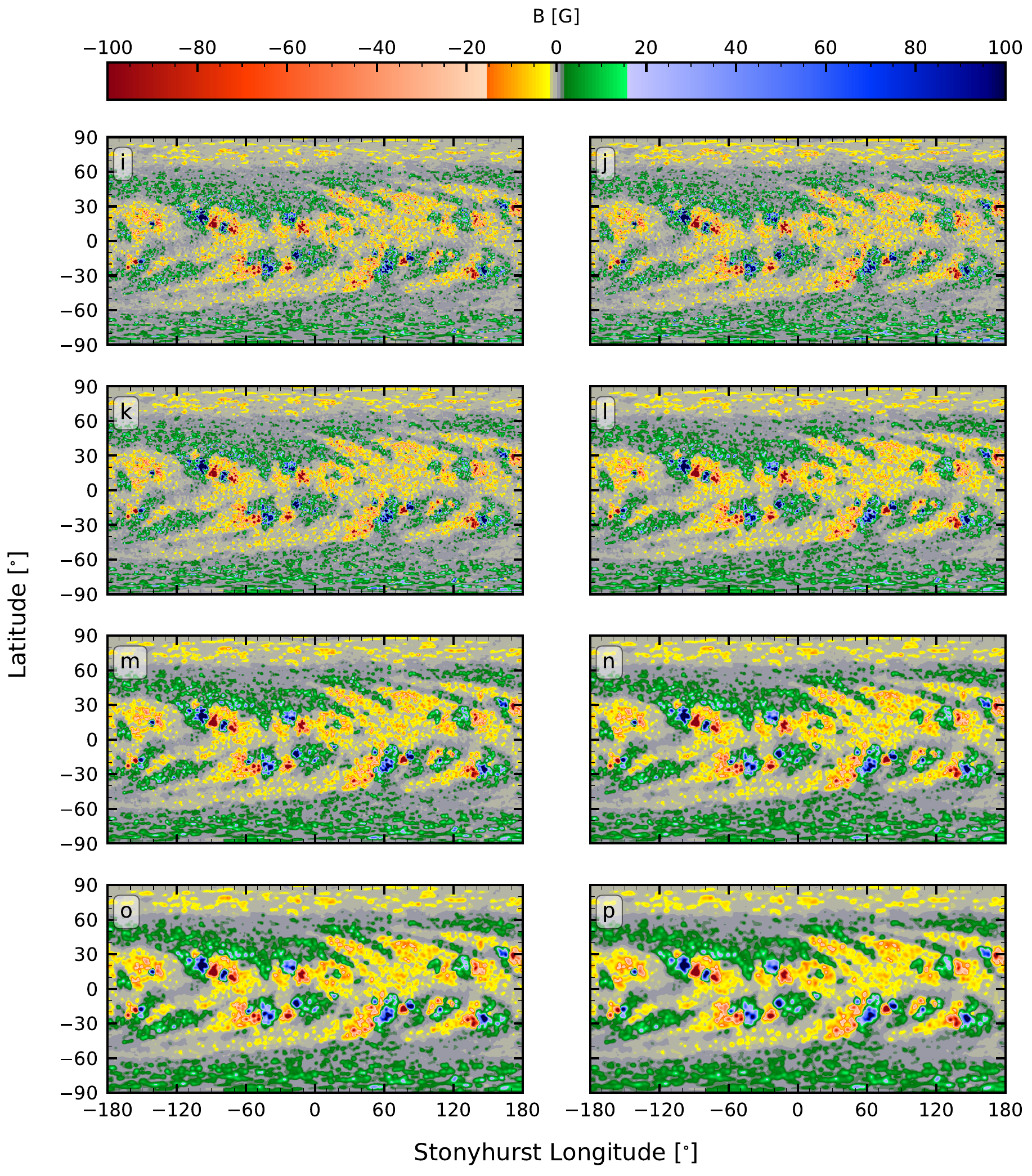}
\caption{Continuation from Figure~\ref{fig:maps1}. (i, j) Synoptic charts with modified polar fields, increased and decreased by $\pm30\%$, respectively. (k-p) Synoptic charts smoothed using Gaussian kernels of varying sizes: 0.5, 1.0, 1.5, 2.0, 2.5, and 3.0 degrees, respectively.} \label{fig:maps2}
\end{figure*}

\section{Methods} \label{sec:methods}
To get a measure of the uncertainties, we run the created set of input magnetograms with three different solar wind prediction models and analyze their variations. In addition, we use the pre-processed \gtm\ with the thermodynamic Magnetohydrodynamic Algorithm outside a Sphere \citep[\mas;][]{MAS1996,Mikic99} model using two different configurations for the heating function in the coronal model part. The \mas\ code solves time-dependent resistive thermodynamic MHD equations in three-dimensional spherical coordinates to derive the plasma and magnetic field of the corona and the ambient solar wind taking into account a realistic energy equation with anisotropic thermal conduction, radiative losses, and coronal heating. The two different heating configurations are referred to as \mas$_{\mathrm{HM2}}$ and \mas$_{\mathrm{WTD}}$, respectively and described in Appendix~\ref{sec:mas}.\\

We run all 16 maps (\textit{i.e.,} the \gtm\ input map plus its 15 different variations) using the Wang-Sheeley-Arge (\wsa) model \citep{Arge2000}, an operational solar wind forecasting tool that combines a potential field source surface (PFSS) model \citep{Altschuler1969} with a Schatten current sheet model \citep[SCS;][]{Schatten1971} to construct the coronal magnetic field. The results are then fed into an empirical solar wind relation to estimate the solar wind at the inner boundary and subsequently propagated to Earth using the 1D kinematic solar wind model within WSA, which simulates the quasi-ballistic propagation of hypothetical solar wind macroparticles through the heliosphere. The same maps are also run with the Heliospheric Upwind eXtrapolation model with time-dependence \citep[\huxt;][]{owens2020, Barnard2022}. \huxt\ is a computationally efficient, open-source solar wind model with reduced physics that we coupled to the coronal output of the \wsa\ model at 21.5 R$_{\odot}$. And lastly we use the European Heliospheric FORecasting Information Asset \citep[\euhforia;][]{Pomoell2018}. It is a three-dimensional MHD model of the inner heliosphere, typically extending to several au that is coupled with a coronal model consisting of a PFSS model and a SCS model extending to the inner boundary of 21.5 R$_{\odot}$. For more details on these models, see Appendix~\ref{sec:modelsetup}. 
\section{Results} \label{sec:res}

\subsection{MHD view of the heliosphere} \label{subs:res_mhd}
\begin{figure}
\centering \includegraphics[width=1\linewidth,angle=0]{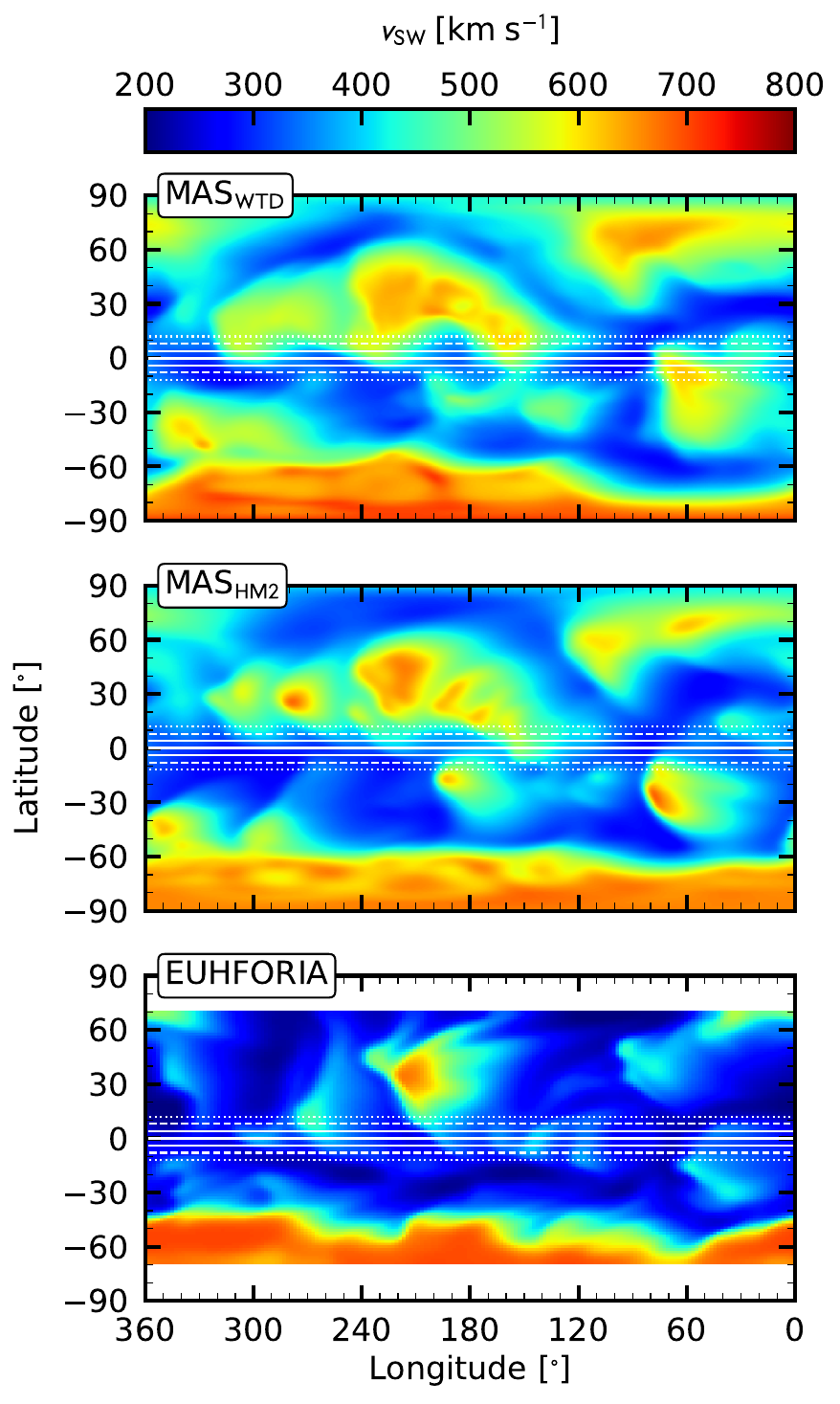}
\caption{Solar wind solutions of the MHD models (from top to bottom: \mas$_{\mathrm{WTD}}$, \mas$_{\mathrm{HM2}}$, and EUHFORIA) for 215R$_{\odot}$ and the processed \gtm\ (\mas) and the \gtm\ for \euhforia. The white solid, dashed and dotted lines correspond $\lambda=0^{\circ}$,$\pm 4^{\circ}$,$\pm 8^{\circ}$,$\pm 12^{\circ}$ respectively. Note that the maps are displayed in counterclockwise longitude to match in-situ solar wind profiles. }\label{fig:mhdmodels}
\end{figure}

We can compare the coronal and heliospheric structures using the \mas\ model results. Our analysis assumes that the \gtm\ accurately represents the concurrent state of the solar magnetic field at any given time. By combining this map with the \mas\ model, which incorporates comprehensive physics, we gain valuable insights. However, due to uncertainties in the setup and initial conditions of MHD models (e.g., heating function and plasma parameters), the \mas\ runs may not necessarily produce a heliosphere that is closer to the ground truth than other models. Therefore, we compare the operational model results with the \mas\ runs to assess the spread of model solutions, and not as a metric of model accuracy. Figure~\ref{fig:mhdmodels} shows the solar wind structure near Earth's orbit at $215~\mathrm{R}_{\odot}$ for the \mas\ runs as well at the \euhforia\ run with the \gtm. The most notable features is a prominent high-speed stream (HSS) extending over at least $100^{\circ}$ in longitude. However, the equatorial cut (indicated by the central white line) does not continuously traverse this region; instead, it intersects the high-speed portion multiple times. Both \mas\ model runs exhibit slight structural differences in the latitude range near Earth's assumed location (presumed as $\lambda=0^{\circ}$), leading to variations in the measured solar wind profile. Sensitivity in the solar wind speed due to slight variations of the target latitude can be seen by using additional latitudinal cuts (at $\lambda=\pm4^{\circ}$,$\pm8^{\circ}$ and $\pm12^{\circ}$). For our \mas\ runs, this results in local changes ($|\Delta v|$) of the solar wind speed of up to $125$~km~s$^{-1}$, $188$~km~s$^{-1}$ and $283$~km~s$^{-1}$ for $\lambda=\pm4^{\circ}$,$\pm8^{\circ}$ and $\pm12^{\circ}$, respectively (see {Figure}~\ref{fig:comparison}). The solar wind profiles are correlated with a \textit{Pearson} correlation coefficient of $cc_{\mathrm{Pearson}} = 0.41$ and a value for the Hanna and Heinold metric of $\mathrm{HH}_{||}=0.23$ (Eq.~11 in Tab.~\ref{tab:metrics}) and the Root Mean Square Error (RMSE, Eq.~4 and~10 in Tab.~\ref{tab:metrics}) between the two profiles is $87.1 \pm 65.3 ~\mathrm{km~s^{-1}}$. This demonstrates that even small changes in the MHD model can significantly impact the measured speed, emphasizing that minor uncertainties in model parameters can result in substantial uncertainties in the outcomes. In the \euhforia\ model, the HSS is significantly smaller in longitudinal and latitudinal extent and is only intersected by $\lambda\ge4^{\circ}$.

\subsection{Uncertainties within the different models}

\begin{figure}
\centering \includegraphics[width=1\linewidth,angle=0]{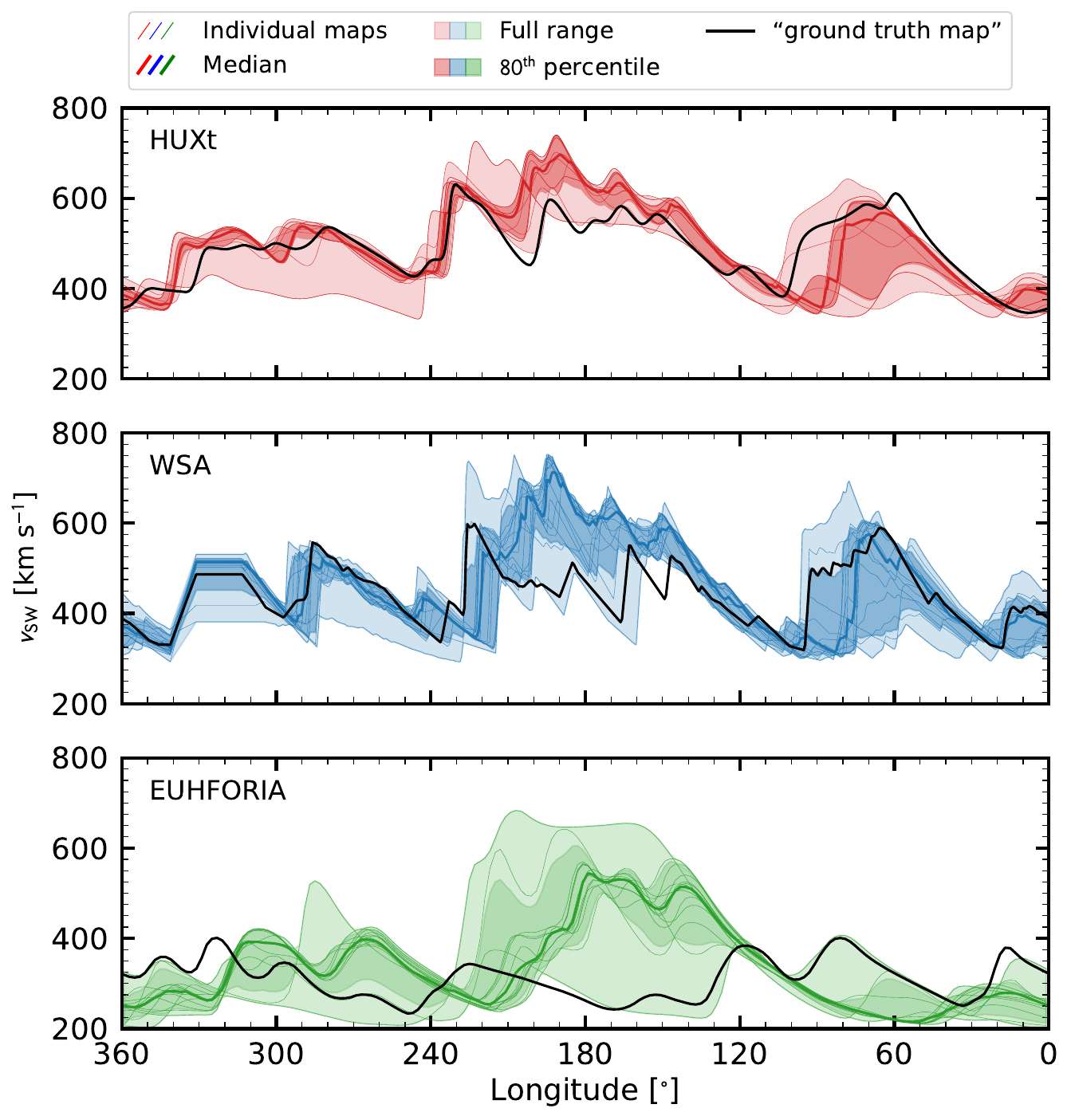}
\caption{Forecast solar wind speeds for \huxt, \wsa\ and \euhforia\ (top to bottom). The black line represents the run with the ground truth map, the thin lines are all the runs using the modified maps, the thick line is the median of all the runs and the shaded areas are the 80th and 100th percentiles. The x-axis represents counterclockwise longitude corresponding to the in-situ solar wind of a rotated steady-state solution at a stationary measurement point.\label{fig:indiv_models}}
\end{figure}

We investigate the uncertainties within each model caused by the different modified input maps to determine the uncertainty range. All 16 maps were used as input to \huxt, \wsa, and \euhforia\ to predict the solar wind speed at Earth's assumed location ($\lambda=0^{\circ}$), as shown in Figure~\ref{fig:indiv_models}. The results generally show a similar structure across all models: a short HSS followed by a extended HSS (approximately $100^{^\circ}$~longitude), and then another shorter one. However, significant differences emerge as not all these parts are capture by all models and maps. The trailing stream specifically is not captured by all runs, although those using the \gtm\ notably all capture it. Furthermore, we observe that the full range of results using the modified maps does not always encompass the outcomes from the \gtm\ runs. \textit{I.e.,} the true solution might not lie within the range of uncertain (modified) maps. Whereas for the \huxt\ runs, the ensemble median corresponds quite well to the \gtm\ run (RMSE is $58.6 ~\mathrm{km~s^{-1}}$, $cc_{\mathrm{Pearson}}=0.75$ and $\mathrm{HH}_{||}=0.12$), for \wsa\ there is a notable difference (RMSE is $86.9 ~\mathrm{km~s^{-1}}$, $cc_{\mathrm{Pearson}}=0.51$ and $\mathrm{HH}_{||}=0.19$), and \euhforia\ even appears to be anti-correlated (RMSE is $120.9 ~\mathrm{km~s^{-1}}$, $cc_{\mathrm{Pearson}}=-0.49$ and $\mathrm{HH}_{||}=0.38$). For more detailed statistics see Table~\ref{tab:stats} in Appendix~\ref{sec:stats}.\\

But as shown in the \mas\ model results, small latitudinal differences can significantly impact the predicted solar wind velocity. To investigate the additional uncertainties arising from this effect, we computed the solar wind speed for all models and maps at additional latitudinal positions ($\lambda=\pm4^{\circ}$, $\pm8^{\circ}$, and $\pm12^{\circ}$). We then determined the deviation of the runs using the \gtm\ map from the corresponding solar wind velocities derived from these latitude intervals. \textit{I.e.}, a larger latitude interval leads to a broader predicted range of solar wind velocities and is, therefore, more likely to encompass the \gtm\ run.\\

\begin{figure}
\centering \includegraphics[width=1\linewidth,angle=0]{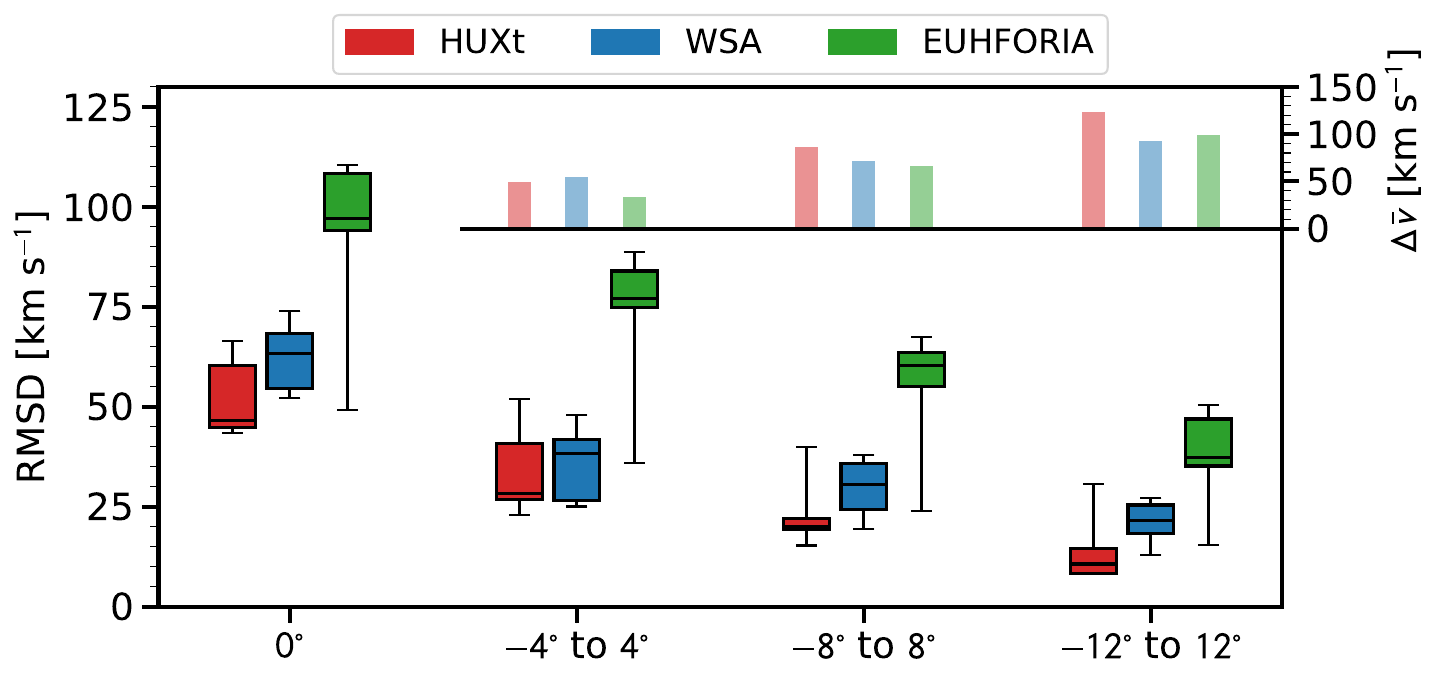}
\caption{Deviation (RMSD) of the runs using the \gtm\ compared to modified runs at $\lambda = 0^{\circ}$ and in the $\pm4^{\circ}$, $\pm8^{\circ}$, and $\pm12^{\circ}$ intervals. The central bar corresponds to the median of all 15 modified map results, the box represents the 20th and 80th percentiles, and the error bars correspond to the 10th and 90th percentiles. The average size of the respective velocity intervals ($\Delta \bar{v}$) is shown in the top right.}\label{fig:RMSD}
\end{figure}

Figure~\ref{fig:RMSD} presents this deviation as the Root Mean Square Deviation (RMSD; see Eq.~12 in Tab.~\ref{tab:metrics}) as a function of the latitude interval. We find that extending the latitude interval to $\lambda=\pm4^{\circ}$, $\pm8^{\circ}$, and $\pm12^{\circ}$ results in average velocity intervals ($\Delta \bar{v}$) of $48$, $76$, and $106~\mathrm{km~s^{-1}}$, respectively. Note that each latitude interval encompasses all solutions within its range. For instance, the $\pm4^{\circ}$ interval includes the solutions at $+4^{\circ}$, $-4^{\circ}$, and $0^{\circ}$.
By doing so, the RMSD (for one profile \textit{e.g.,} at  $\lambda=0^{\circ}$ the RMSD defaults to the RMSE) decreases by $20$–$40\%$, $38$–$57\%$, and $61$–$77\%$ for the different latitude intervals. This demonstrates that expanding the latitude range around the proposed Earth's position can significantly improve the accuracy of forecasts by increasing the likelihood that the ground truth lies within the predicted interval.\\

For the individual maps, we find little difference in the uncertainties, except for the significant increase in uncertainty when comparing maps based on the \gtm\ to those based on the synoptic chart approximation. The RMSE increases by at least $10$, $40$, and $70~\mathrm{km~s^{-1}}$ for \huxt, \wsa, and \euhforia, respectively.

\subsection{Uncertainties in comparison to \mas\  model runs}

\begin{figure*}
\centering \includegraphics[width=1\linewidth,angle=0]{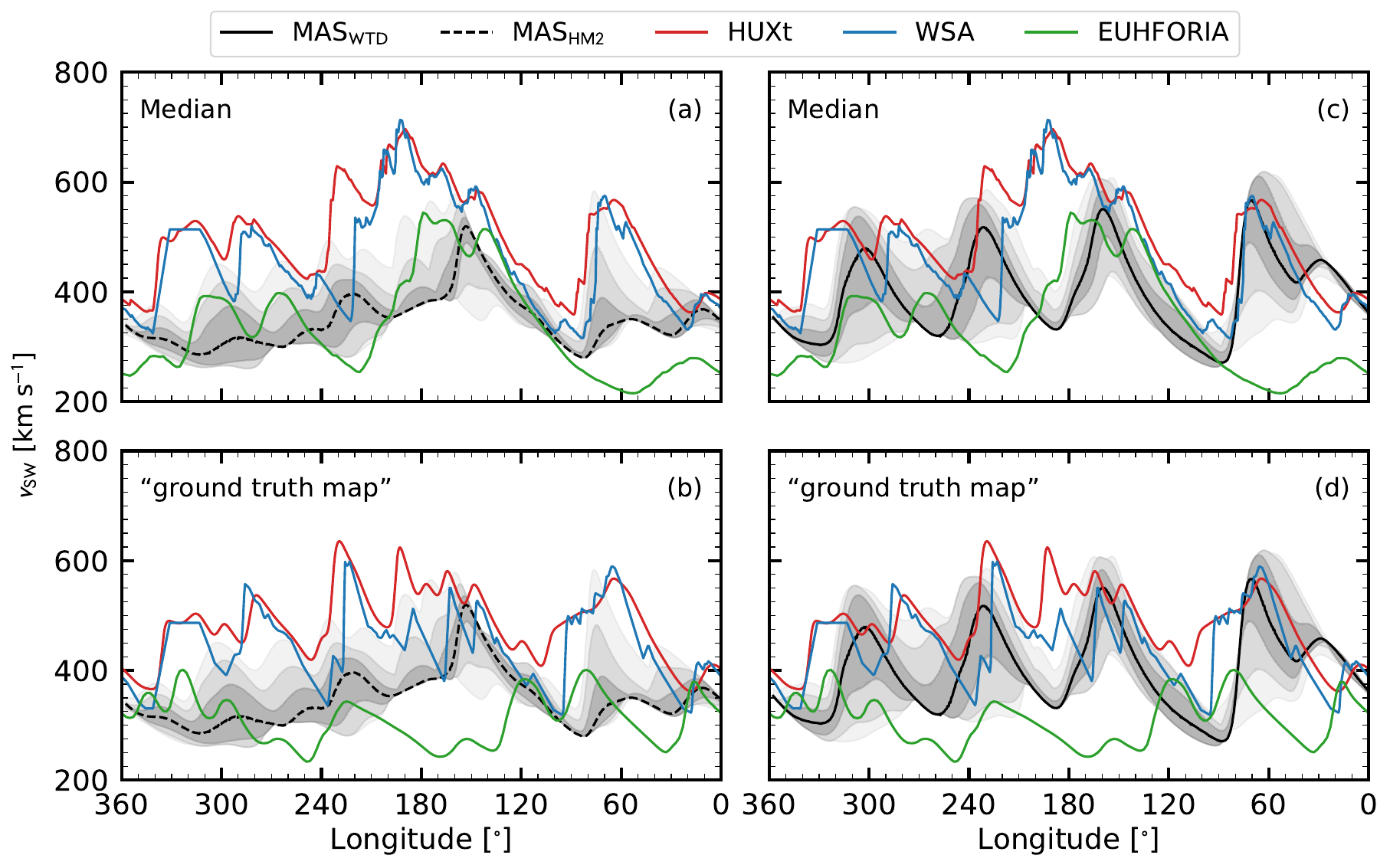}
\caption{Comparison of the \mas\ results with the results of the full sample of runs using the modified and "ground truth" maps. Panels (a) and (b) show the \mas$_{\mathrm{WTD}}$ results, while panels (c) and (d) show the \mas$_{\mathrm{HM2}}$ results. The solid and dashed lines represent the cut at $\lambda = 0^{\circ}$ through the steady-state solution, and the shaded areas represent the solar wind in $\lambda = \pm4^{\circ}$, $\pm8^{\circ}$, and $\pm12^{\circ}$ intervals, respectively.}\label{fig:comparison}
\end{figure*}

To better understand the uncertainty ranges when incorporating different models, we compare the solar wind predictions from \huxt, \wsa, and \euhforia\ using both the sample median and the results from the \gtm\ with the solar wind predictions from the \mas\ model runs. This comparison is shown in Figure~\ref{fig:comparison}. We find that the predictions from the three operational models differ significantly from those of the \mas\ runs, both in terms of the median of the modified runs and the \gtm\ results. While we find that the \huxt\ and \wsa\ results for the respective sample median and \gtm\ run are in good agreement ($\mathrm{HH}_{||} = 0.13$ and $0.14$ and $cc_{\mathrm{Pearson}} =0.73$ and $0.87$), the \euhforia runs deviate from both ($0.41 \leq \mathrm{HH}_{||}  \leq 0.51$).  However, most of these runs are significantly different from the \mas\ runs. This discrepancy is reflected in the spread of values of $\mathrm{HH}{||}$, ranging from $0.06$ to $0.41$, with no clear correlation in most cases. The average RMSE against \mas$_{\mathrm{WTD}}$ solar wind solutions is approximately $121~\mathrm{km~s^{-1}}$, ranging from $95$ to $135~\mathrm{km~s^{-1}}$. In contrast, RMSE values for the three models compared to \mas$_{\mathrm{HM2}}$ results vary more widely, from $77$ to $172~ \mathrm{km~s^{-1}}$, with a mean of $127~\mathrm{km~s^{-1}}$. And even when considering different latitudinal ranges in the \mas\ model (gray shaded areas in {Figure}~\ref{fig:comparison}), the agreement does not significantly improve. Further statistics are listed in Table~\ref{tab:stats} in Appendix~\ref{sec:stats}.

\section{Discussion and Conclusion} \label{sec:disc}
From a solar rotation of synthetic magnetograms, we created a set of modified maps designed to mimic uncertainties in the magnetic field maps caused by incomplete observational constraints. Using this set of input maps, we ran multiple operational solar wind forecasting models to derive a plausible range of values for the forecast and, subsequently, estimated the uncertainties.\\

Constraining the uncertainties in solar wind predictions have become more important in recent times. \cite{Bertello2014} concluded that the uncertainties that arise from creating synoptic charts have a significant impact on the location and structure of the neutral lines and the distribution of the coronal holes, which is paramount for predicting the solar wind at a single point in space, \textit{i.e.} Earth. This effect can be seen in the runs using the \gtm\ and \mas\ runs where the large central high-speed stream is not traversed continuously. \cite{Poduval2020} estimated an uncertainty with a RMSE of $85-110~$km$~$s$^{-1}$ in solar wind predictions due to the uncertainties in the photospheric flux density synoptic maps. This is similar to the results derived in this study of $59-121~$km$~$s$^{-1}$ for within the individual models and about $77-172~$km$~$s$^{-1}$ in comparison to the thermodynamic \mas\  models using the \gtm. Further, we show that by using a cloud of points in latitude around the target latitude can significantly reduce the uncertainties. \\

In recent studies, the uncertainties in solar wind forecasting compared to observations for operational models were found to have a RMSE of $100$–$120~$km$~$s$^{-1}$ \citep[\textit{e.g.} see][]{Reiss2016,Reiss2020,milosic2023}. This value is similar to the results obtained within a single model in comparison to when using the \gtm\ maps as input. However, it is important to note that our analysis only compared model runs with an assumed \gtm\ and its modifications. We did not account for the fact that none of these models can fully describe the coronal and heliospheric processes.\\

In summary, we can conclude the following:

\begin{itemize}
    \item For each operational model, we find significant variations between the \gtm\ run and the modified map runs. The RMSE values within \huxt, \wsa, and \euhforia\ are $59~\mathrm{kms^{-1}}$, $87~\mathrm{kms^{-1}}$, and $121~\mathrm{km~s^{-1}}$, respectively.

    \item We find that the largest source of uncertainty stems from the generation of synoptic charts, with the assumption that this is due to the combined impact of the 'aging effect' and the absence of far-side information.

    \item For all three models, the \gtm\ run does not always fall within the range of modified runs, underscoring that ensemble predictions may not fully capture real solar wind variability. However, considering multiple latitudes rather than a single point reduces forecast uncertainty and lowers the chance of excluding the \gtm\ run by $20$–$77\%$.

    \item While the \huxt\ and \wsa\ results are rather similar, the average spread of all the model results is ${\Delta \bar v~=251~\mathrm{km~s}^{-1}}$.

    \item On average, the variations resulting from modifications to the magnetic field input maps, used as a proxy for observational uncertainties, are generally consistent with those found in other studies. Furthermore, the uncertainties within a single model are smaller than the differences between all the models.
\end{itemize}

While it is possible to quantify the uncertainties arising from incomplete magnetic field information by using a \gtm\ as a reference, the true solar wind remains elusive. The processes from solar wind acceleration to its propagation in the heliosphere are not yet fully understood \citep[][]{Viall2020}. Our results suggest that increasing the operational coverage of concurrent solar magnetic field observations could reduce the uncertainty in the background solar wind by approximately $100~\mathrm{km~s^{-1}}$ near 1au. Alternatively, by predicting a range of velocities arising from a cloud of points around Earth can help mitigate some uncertainties arising from incomplete magnetic field information.

\begin{acknowledgments}
We thank the International Space Science Institute (ISSI, Bern) for the generous support of the ISSI teams “Magnetic open flux and solar wind structuring in heliospheric space” (2019--2021) and “Quantitative Comparisons of Solar Surface Flux Transport Models” (2024--2025). This research was funded in whole, or in part, by the Austrian Science Fund (FWF) Erwin-Schr\"odinger fellowship J-4560. Open access funded by Helsinki University Library. SGH thanks Andreas Wagner for valuable discussions. JP and SGH acknowledge funding from the Academy of Finland project SWATCH (343581). Work at Predictive Science Inc. was supported by the NASA LWS Strategic Capabilities Program (grant 80NSSC22K0893). MO  was part-funded by Science and Technology Facilities Council (STFC) grant number ST/V000497/1 and Natural Environment Research Council (NERC) grant NE/Y001052/1.  LAU and BKJ were supported by the NASA LWS Strategic Capabilities Program (grant 80NSSC22K0893) and by NASA grant NNH18ZDA001N-DRIVE to the COFFIES DRIVE Center, managed by Stanford University.

\end{acknowledgments}

\vspace{5mm}

\appendix

%\section{Magnetograms}\label{sec:mags}
%In this Appendix, we present the various magnetic maps utilized in this study, as shown in Figure~\ref{fig:maps1} and Figure~\ref{fig:maps2}.

\section{Model Setup} \label{sec:modelsetup}
In this Appendix, we provide additional details about each of the models that were used for the analysis described in the paper (e.g., \texttt{AFT}, \mas, \wsa, and  \huxt). 

\subsection{Advective Flux Transport model \texttt{AFT}} \label{sec:aft}

The Advective Flux Transport (\texttt{AFT}) model solves the horizontal components of the induction equation, similar to other Surface Flux Transport (SFT) models. However, unlike traditional SFT models, \texttt{AFT} employs convective simulations of horizontal surface flows instead of the diffusivity parameter typically used in other models. These convective simulations include both axisymmetric flows (differential rotation and meridional flow) and convective flows \citep{Conflow2010}. By incorporating convective simulations, \texttt{AFT} eliminates the need for an explicit diffusion term. However, diffusion is still solved within the model solely to stabilize the numerical scheme \citep{AFT2014, AFT2014b}. 

To create synthetic full-Sun magnetic data, \texttt{AFT} utilizes the Synthetic Active Region Generator \citep[\texttt{SARG};][]{Jha2024}, which is incorporated as the source term along with random flux. \texttt{SARG} uses ISSNv2.0 to define the monthly sunspot number (or cycle amplitude) as a function of time. This determines the frequency of active region emergence \citep[see details][]{Jha2024}. \texttt{SARG} uses the known statistical properties of solar active regions, such as the location, flux, and polarity (Joy's Law, Hale's Polarity Law),  to define these paramater for the synthetic active region. The output is a synthetic active region catalog consisting of latitude, longitude, and flux for each bipole of the active regions. Once all these properties are determined, we generate idealized bipoler Gaussian spots. These spots are added into the model, thus simulating the emergence of magnetic active regions. We note that each active region generated by \texttt{SARG} is added to the model only once at the time of its emergence in \texttt{SARG}.

In this particular study \texttt{AFT} also employs the Gaussian random flux generator (GRaFg), with a mean of zero and a standard deviation of one, to emulate the random emergence of small-scale flux. The random flux values drawn from Gaussian distribution are inserted at random locations, where latitude values are sampled from a cosine distribution and longitude values from a uniform distribution. The GRaFg makes sure that the net flux added into the model is always zero to avoid accumulation of any monopole. The number of random grid points and the frequency of emergence of these random fluxes are optimized to ensure that the total absolute flux in the model remains consistent with the case where no random flux is present. We note that in \texttt{AFT}, these random fluxes have no impact on the global properties of the system. Here, we used the GRaFg model with a mean of zero and a standard deviation of one, adding random flux values at 1,000 grid points every hour.

\subsection{Magnetohydrodynamic Algorithm outside a Sphere model \mas} \label{sec:mas}

The Magnetohydrodynamic Algorithm outside a Sphere (\mas) code is an in-production MHD model with over 25 years of ongoing development used extensively in solar physics research \cite{mikic99a,2003PhPl...10.1971L,lionello06a,2009ApJ...690..902L,2011ApJ...731..110L,2013Sci...340.1196D,ECLIPSE2017,lionelloetal2013,torok18}\footnote{\url{https://www.predsci.com/mas}}.  The code is included in the Corona-Heliosphere (CORHEL) software suite \cite{rileyetal2012} hosted at NASA's Community Coordinated Modeling Center (CCMC)\footnote{\url{https://ccmc.gsfc.nasa.gov}} allowing users to generate quasi-steady-state MHD solutions of the corona and heliosphere, as well as the CORHEL-CME suite \citep{Linker2024} which allows users to run MHD simulations of {coronal mass ejections (CME)} events from the Sun to Earth \cite{Linker_2024}\footnote{\url{https://ccmc.gsfc.nasa.gov/models/CORHEL-CME~1}}.
\mas\  is written in modern Fortran and parallelized to run across multiple multi-CPU and multi-GPU systems \citep{Caplan2023}.  {The \mas\ code employs a logically rectangular, non-uniform staggered spherical grid and applies finite-difference discretization techniques, utilizing a combination of explicit and implicit time-stepping schemes.}

MAS contains numerous coronal heating models, both empirical and physics-based, that can be combined together. Empirical heating can be set by a combination of analytic functions \citep{Lionelloetal2009}.  A form of this model that is used in CORHEL is denoted in this work as {\tt HM2} \citep{2010AGUFMSH42A..09M}.  The primary physics-based heating approach in MAS is the wave-turbulence-driven (WTD) model \citep{lionello2014,downs2016,Mikicetal2018} and is denoted as {\tt WTD} in this work.

\subsection{Wang-Sheeley-Arge model \wsa} \label{sec:wsa}
The \wsa model consists of three components. The first, a PFSS model, maps the inner coronal magnetic field from the photosphere to source surface radius (R$_{\mathrm{SS}}$) using a spherical harmonic expansion to match photospheric flux and enforce radial field at the source surface. The second component extends the field solution from the SCS radius (R$_{\mathrm{SCS}}$) to 5 solar radii using the SCS model. This approach creates a current sheet and helmet streamer above the neutral line by temporarily enforcing positive radial field values and solving a potential field problem, ensuring open field lines remain open and minimizing energy. The final component uses the field at 5 solar radii to derive the wind speed, based on an empirical formula that considers the local magnetic flux expansion, $f_{exp}$, and the distance to the nearest coronal hole boundary, $\Theta_B$.\\

For the \wsa performed in this study, we used R$_{\mathrm{SS}} =2.51$R$_{\odot}$, R$_{\mathrm{ScS}} =2.49$R$_{\odot}$ as well as the empirical solar wind relation:
\begin{equation}
    V = V_0 + \dfrac{V_m}{(1+f_{\mbox{\scriptsize exp}})^{C_1}}\,\left(1 - C_2\,\mbox{exp}\left[-\left(\dfrac{\Theta_B}{C_3}\right)^{C_4}\right]\right)^{C_5}
    \label{eqn:wsa_windspeed}
\end{equation}
where $V_0$, $V_m$, and $C_1$ - $C_5$ are empirically determined parameters given in Table \ref{table:wsa_parameters}.
\begin{table}
%  \centering
\qquad \qquad \,
  \begin{tabular}{|c|c|c|c|c|c|c|}
    \hline
    $V_0$ & $V_m$ & $C_1$ & $C_2$ & $C_3$ & $C_4$ & $C_5$ \\
    \hline 
    286 m/s& 625 m/s& 2/9 & 0.8 & $1$ & 2 & 3 \\
    \hline
  \end{tabular}
  \caption{Parameter values used for solar wind speed predictions in Equation \ref{eqn:wsa_windspeed}.}
  \label{table:wsa_parameters}
\end{table}

\subsection{Heliospheric Upwind eXtrapolation with Time-dependence model \huxt} \label{sec:huxt}

The Heliospheric Upwind eXatrpolation model with time dependence \citep[HUXt][]{owens2020, Barnard2022} is a numerical model that approximates the solar wind as a 1-dimensional, incompressible hydrodynamic flow \citep[see also ][]{Riley2011a}. Consequently, solar-wind flows are dynamic and may accelerate and decelerate through stream interaction, but only solar wind speed, (and not other plasma parameters) is reconstructed. Despite the  approximations employed, HUXt produces solar-wind speed structure throughout the model domain (typically from 0.1 to 1.5~AU) that agrees very closely with the results of 3D MHD models for the same boundary conditions \citep{owens2020}, but at a fraction of the computational cost. This allows for large parameteric studies, and for ensemble forecasting and data assimilation techniques to be more easily employed.

\subsection{EUropean Heliospheric FOrecasting Information Asset \euhforia} \label{sec:euhforia}

%\textcolor{red}{@JENS can you please put in some text here, model setup etc? thank you}

\euhforia is a framework for modeling global-scale dynamics in the inner heliosphere. Given information of the plasma and magnetic field at the heliospheric inner boundary set at a heliocentric radius of $r_H = 0.1$ au, the evolution of the solar wind for $r > r_H$ is determined by solving the magnetohydrodynamic (MHD) equations in three dimensions, as detailed in \citet{Pomoell2018}.

The ambient state of the solar wind at $r=r_H$ can be prescribed in various ways. Commonly and for the runs used in this work, it is constructed using a semi-empirical prescription akin to the approach employed by the \wsa model (see \ref{sec:wsa}). First, the magnetic field in the coronal domain ($r \in [R_\odot, r_H])$ is modeled using the PFSS model in the lower corona and a current sheet model in the upper corona. In this work, these models are computed using a finite-difference approach, and with the source surface radius in the PFSS model set to $2.51 R_\odot$ and the inner radial boundary of the current sheet model set at $2.49 R_\odot$. The choice of these parameters can have a significant influence on the resulting magnetic field structure \citep[e.g.,][]{Asvestari2019}, and were here chosen to be the same as used by the \wsa model. Before computation, the (numerical) signed flux in the considered magnetic maps was balanced to zero using a multiplicative method. The computation employed a constant radial grid spacing of $\Delta r \sim 0.005 R_\odot$ in the PFSS model and $\Delta r \sim 0.05 R_\odot$ in the current sheet model, and a $1^\circ$ angular grid spacing as given by the magnetic maps. 

Using the coronal model, the connectivity of the magnetic field is determined by tracing field lines, and a map of the (great circle) distance $d$ to nearest open field regions constructed. The wind speed is then determined using the empirical prescription given by
\begin{equation}
    V = V_\mathrm{s} + \frac{1}{2}(V_\mathrm{f} - V_\mathrm{s}) \left( 1 + \tanh \left( \frac{d - d_0}{w} \right) \right)
\end{equation}
where $V_\mathrm{f} = 720$ km/s, $V_\mathrm{s} = 220$ km/s, $d_0 = 2^\circ$ and $w = 2^\circ$. 
A similar empirical relation was studied in \citet{Rileyetal2015}. 
Based on the obtained empirical solar wind speed, the boundary data for the MHD computation is determined as described in \citet{Pomoell2018}, with the following modifications: First, the the absolute value of the radial magnetic field at $r_H$ is constant and chosen to provide an equal unsigned flux as the unsigned open flux given by the coronal model while the field polarity is given directly by the coronal model, and second, the wind speed is reduced not by a constant but instead by using $v_r = V - (40 \, \mathrm{km/s}) \frac{600 \, \mathrm{km/s}}{V}$.  

The heliospheric MHD computation is run for a duration of 10 days until an approximate steady-state solar wind solution is reached (no CMEs were launched). The simulation domain encompassed a radial domain $r \in [0.1, 1.5]$ au and a latitudinal domain $\lambda \in [-70^\circ, 70^\circ]$. A uniform grid using $(300, 93, 180)$ cells in the radial, latitudinal and longitudinal directions was employed.

\section{Statistics} \label{sec:stats}
Table~\ref{tab:metrics} lists the metrics used in this study and Table~\ref{tab:stats} shows the statistics of the runs using the modified maps to those using the \gtm.\\

\begin{table}[h!]
    \centering
    \renewcommand{\arraystretch}{1.3}
    \begin{tabular}{|l|l|c|c|}
        \hline
        \textbf{Nr.} &\textbf{Name} & \textbf{Abbrev.} & \textbf{Equation}   \\
        \hline
        1 & Mean Absolute Deviation & MAD & $\text{MAD} = \frac{1}{n} \sum_{i=1}^{n} \left| M_i - T_i \right|$  \\
        \hline
        2 & Mean Error & ME & $\text{ME} = \frac{1}{n} \sum_{i=1}^{n} \left( M_i - T_i \right)$   \\
        \hline
        3 & Mean Squared Error & MSE & $\text{MSE} = \frac{1}{n} \sum_{i=1}^{n} \left( M_i - T_i \right)^2$   \\
        \hline
        4 & Root Mean Squared Error & RMSE & $\text{RMSE} = \sqrt{\text{MSE}}$   \\
        \hline
        5 & Mean Percentage Error & MPE & $\text{MPE} = \frac{1}{n} \sum_{i=1}^{n} \frac{M_i - T_i}{T_i} \times 100$  \\
        \hline
        6 & Std. Dev. of MAD & $\sigma_{\mathrm{MAD}}$ & $\sigma_{\mathrm{MAD}} = \sqrt{\frac{1}{n-1} \sum_{i=1}^{n} \left( |M_i - T_i| - \text{MAD} \right)^2}$   \\
        \hline
        7 & Std. Dev. of ME & $\sigma_{\mathrm{ME}}$ & $\sigma_{\mathrm{ME}} = \sqrt{\frac{1}{n-1} \sum_{i=1}^{n} \left( (M_i - T_i) - \text{ME} \right)^2}$   \\
        \hline
        8 &Std. Dev. of MSE & $\sigma_{\mathrm{MSE}}$ & $\sigma_{\mathrm{MSE}} = \sqrt{\frac{1}{n-1} \sum_{i=1}^{n} \left( (M_i - T_i)^2 - \text{MSE} \right)^2}$   \\
        \hline
        9 & Std. Dev. of MPE & $\sigma_{\mathrm{MPE}}$ & $\sigma_{\mathrm{MPE}} = \sqrt{\frac{1}{n-1} \sum_{i=1}^{n} \left( \frac{M_i - T_i}{T_i} \times 100 - \text{MPE} \right)^2}$   \\
        \hline
        10$^{*}$ & Std. Dev. of RMSE & $\sigma_{\mathrm{RMSE}}$ & $\sigma_{\mathrm{RMSE}} \approx \frac{1}{2 \sqrt{\text{MSE}}} \cdot \sigma_{\mathrm{MSE}}$  \\
        \hline
        11$^{**}$ & Hanna and Heinold metric & $\mathrm{HH}_{||}$ & $\mathrm{HH}_{||} = \sqrt{\dfrac{\sum_{i=1}^n\,|M_i-T_i|^2}{\sum_{i=1}^n\,|M_i||T_i|}}$  \\
        \hline
       12$^{***}$ & Root Mean Square Deviation & RMSD & RMSD $ =
        \begin{cases}
        0, & \text{if } L_i \leq T_i \leq U_i \\
        \sqrt{\frac{1}{n} \sum_{i=1}^{n} (L_i - T_i)^2}, & \text{if } T_i < L_i \\
        \sqrt{\frac{1}{n} \sum_{i=1}^{n} (U_i - T_i)^2}, & \text{if } T_i >  U_i \\
        \end{cases}$\\ 
        \hline
    \end{tabular}
    \vspace{0.3cm} \\
    \begin{minipage}{0.95\textwidth}
        \small
        $M_i$: Value of runs using modified maps (or median value). \newline
        $T_i$: Value of run using the \textit{ground truth model} (\textsc{GTM}). \newline
        $n$: Total number of data points. \newline
        $L_i$: Lower bound. \newline
        $U_i$: Upper bound. \newline
        $^{*}$: Approximated using delta method. \newline
        $^{**}$: We are calculating an absolute value version of the Hanna and Heinold metric \citep{HHbook} because it provides a normalized, scale-invariant measure of similarity between two time series. By accounting for both the differences and the magnitudes of the data points, this metric allows us to compare the relative shapes and patterns of the series regardless of their absolute scale. This error metric has been demonstrated to be more effective for evaluating numerical solutions compared to the more commonly used normalized root mean square (RMSE) error metric \citep{HHgtNRMSD}. \newline
        $^{***}$: Given two data series representing the lower and upper bounds, RMSD quantifies the deviation of a comparison data series from this interval. If the lower and upper bounds are identical RMSD defaults to RMSE.
    \end{minipage}
    \caption{Summary of statistical metrics used.}
    \label{tab:metrics}
\end{table}

\begin{table}[ht]
    \centering
    \caption{Statistics of the deviation of the median of the runs using the modified maps and those using the \gtm\ against the two runs using the \mas\  model (\mas$_{\mathrm{HM2}}$ and \mas$_{\mathrm{WTD}}$). From left to right: Mean Absolute Error (MAD) and the standard deviation of MAD ($\sigma_{\mathrm{MAD}}$), the Mean Square Error (MSE) and the standard deviation of the MSE ($\sigma_{\mathrm{MSE}}$), the Root Mean Square Error (RMSE) and the error in the RMSE ($\sigma_{\mathrm{RMSE}}$), the maximum deviation (D$_{\mathrm{max}}$) and the absolute Hanna and Heinold metric ($\mathrm{HH}_{||}$).}
    \begin{tabular}{|c|ccccccccccc|}
        \hline
        \multirow{2}{*}{\textbf{Model}} & MAD & $\sigma_{\mathrm{MAD}}$ & ME & $\sigma_{\mathrm{ME}}$ & RMSE & $\sigma_{\mathrm{RMSE}}$  & MPE & $\sigma_{\mathrm{MPE}}$ &  D$_{max}$ & $\mathrm{HH}_{||}$ & $cc_{\mathrm{Pearson}}$\\
        & [km s$^{-1}$] & [km s$^{-1}$] & [km s$^{-1}$] & [km s$^{-1}$] & [km s$^{-1}$] & [km s$^{-1}$] & [$\%$] & [$\%$] &  [km s$^{-1}$] & [ ] & [ ]\\
        \hline \hline

 & \multicolumn{11}{c|}{\textbf{Median of modified runs vs. \gtm\ runs}}  \\ \hline
        \huxt & 41.8 & 42.0 & 10.7 & 58.3 & 59.2 & 58.6 & 2.4 & 11.7 & 203.1 & 0.12 & 0.75 \\
        \hline
        \wsa & 56.5 & 66.1 & 17.2 & 85.3 & 86.9 & 84.8 & 4.4 & 18.3 & 268.6 & 0.19 & 0.51\\
        \hline
        \euhforia & 94.8 & 75.2 & 25.1 & 118.5 & 120.9 & 88.6 & 12.4 & 42.4 & 287.3 & 0.38 &-0.49\\ 
        \hline \hline

& \multicolumn{11}{c|}{\textbf{Median of modified runs vs. MAS$_{\mathrm{HM2}}$}}  \\ \hline
\wsa & 114.1 & 90.6 & 112.8 & 92.3 & 145.7 & 90.5 & 33.6 & 27.8 & 359.9 & 0.36 & 0.35 \\ \hline
\huxt & 151.8 & 80.5 & 151.8 & 80.5 & 171.8 & 75.6 & 44.8 & 24.4 & 332.8 & 0.41 & 0.42 \\ \hline
\euhforia & 64.4 & 42.3 & -10.2 & 76.5 & 77.1 & 40.8 & -2.8 & 21.9 & 163.8 & 0.22 & 0.53 \\ \hline
 & \multicolumn{11}{c|}{\textbf{Runs with the \gtm vs. MAS$_{\mathrm{HM2}}$}}  \\ \hline
\wsa & 98.7 & 76.8 & 96.3 & 79.8 & 125.0 & 73.7 & 29.9 & 25.9 & 269.5 & 0.32 & 0.06 \\ \hline
\huxt & 142.3 & 74.5 & 141.3 & 76.5 & 160.6 & 66.2 & 42.7 & 25.1 & 272.9 & 0.39 & 0.21 \\ \hline
\euhforia & 63.3 & 55.2 & -36.1 & 75.8 & 83.9 & 72.1 & -8.3 & 19.9 & 244.6 & 0.26 & -0.33 \\ \hline
 & \multicolumn{11}{c|}{\textbf{Median of modified runs vs. MAS$_{\mathrm{WTD}}$}}  \\ \hline
\wsa & 91.8 & 82.3 & 61.2 & 107.0 & 123.2 & 101.6 & 18.2 & 29.2 & 354.6 & 0.29 & 0.24 \\ \hline
\huxt & 109.8 & 78.8 & 100.2 & 90.7 & 135.1 & 96.4 & 27.9 & 26.2 & 363.4 & 0.30 & 0.39 \\ \hline
\euhforia & 101.6 & 80.3 & -61.8 & 113.8 & 129.4 & 92.0 & -12.8 & 25.6 & 333.6 & 0.35 & 0.06 \\ \hline
 & \multicolumn{11}{c|}{\textbf{Runs with the \gtm vs. MAS$_{\mathrm{WTD}}$}}  \\ \hline
\wsa & 72.1 & 61.5 & 44.7 & 83.6 & 94.7 & 69.6 & 14.4 & 25.2 & 246.6 & 0.22 & 0.30 \\ \hline
\huxt & 99.3 & 63.5 & 89.7 & 76.5 & 117.9 & 71.1 & 25.5 & 24.4 & 272.0 & 0.27 & 0.44 \\ \hline
\euhforia & 109.7 & 67.0 & -87.7 & 94.0 & 128.5 & 69.8 & -18.7 & 21.9 & 292.7 & 0.37 & -0.20 \\ \hline

    \end{tabular}
    \label{tab:stats}
\end{table}

%\bibliography{bib}{}
\bibliographystyle{aasjournal}

\end{document}